June 18, 2010

# A New Principle in Physics: the Principle of "Finiteness", and Some Consequences

Abraham Sternlieb
Princeton University-Plasma Physics Laboratory
P.O. Box 451, Princeton, New Jersey 08543

## Abstract

In this paper I propose a new principle in physics: the principle of "finiteness". It stems from the definition of physics as a science that deals (among other things) with measurable dimensional physical quantities. Since measurement results, including their errors, are always finite, the principle of finiteness postulates that the mathematical formulation of "legitimate" laws of physics should prevent exactly zero or infinite solutions. Some consequences of the principle of finiteness are discussed, in general, and then more specifically in the fields of special relativity, quantum mechanics, and quantum gravity. The consequences are derived independently of any other theory or principle in physics. I propose "finiteness" as a postulate (like the constancy of the speed of light in vacuum, "c"), as opposed to a notion whose validity has to be corroborated by, or derived theoretically or experimentally from other facts, theories, or principles.

## 1. Introduction: the Principle of "Finiteness"

The notion of finiteness of physical entities appears in the scientific literature for quite some time, as being opposed to either the notion of absolute zero or to the notion of infinity. The Born and Infeld model [1], trying to deal with the infinite self-energy of a point charge, was founded on a "principle of finiteness" that says (in their words) that a satisfactory theory should avoid physical quantities becoming infinite (this was before the arrival of renormalization). Albert Einstein [2], referring to the meaning of $R_s$ (Schwarzschild radius) and the black-hole singularity, wrote that physical models, cannot exhibit such a singularity, including the need to modify his own general relativity in the future, when new relevant evidence will be available.

A. Vilenkin [3] proposes a cosmological model that is intended to avoid the Big-Bang singularity.

T. D. Lee [4] mentions the necessity of time discretization because of the finite number of measurements needed to be performed in a finite time.

Gambini and Pullin [5] discuss loss of coherence and unitarity in the measurement process and consequently obtain an expression for a minimum possible clock accuracy:

$$\delta t \propto t_{pl}^{2/3} \bullet t^{1/3} \qquad (t_{pl} = \text{Planck time}).$$

Refs. [6] and [7] mention that infinities generally arise because of the point-like definition of elementary particles, and that string theory introduces finiteness (finite-sized strings) as a possible remedy. Finally, to conclude this literature sampling, in Ref. [8]

mention is made of the (until now unknown) contributions of Ettore Majorana to the ideas of elementary (finite) length and time scales. The central theme evolving from the above-mentioned references, is that there is no physical reality besides measureable entities. In other words, there is no place in physical models for zero-size (point-like) entities or for infinite (singular) entities.

It is clear that the measurement process and its outcome are of outmost importance in defining the "legitimate" content of physics, as we shall see in the following. We begin our discussion by mentioning summarily the axiomatic basis of special relativity.

Albert Einstein developed his special relativity theory on the basis of two principles (postulates): a) the constancy of the speed of light, and b) the principle of special relativity, that requires invariance of the mathematical expressions of "legitimate" physical laws (under appropriate transformations) when passing from one inertial frame of reference to another which is moving at a constant velocity with respect to the first. The two postulates were chosen because of being based on experimental evidence and on its logical consequences, as available at that time.

By adhering to his postulates and studying carefully the measurement process, Einstein derived the Lorentz Transformations (which had been invented before him, ad-hoc, by Lorentz), which explained naturally the length contraction and time dilation. As a consequence, the immovable ether frame of reference became superfluous.

New principles (postulates) in physics do not appear frequently. They should be based on well-established experimental facts and/or on reasonable logical thinking. Also, they should be as general as possible, in order to be relevant to most of physics (Special and General Relativity, Quantum Mechanics, etc.)

Accordingly, it seems desirable to derive such a postulate from the very definition of physics. This would be a fundamental postulate, to which the laws of physics should obey.

I choose to (ad-hoc) define physics as follows: physics is the science that deals (not exclusively) with well-defined physical quantities, including theories about their functional relationships. It also deals with measurement methodologies relevant to the measurable quantities. Concerning the complex relationship between theory and measurement (experiment), one should consult books on the philosophy of science [9, 10], which is outside the scope of this paper.

In measurement methodology there is always a need to include a measurement error ("accuracy", or "error bar") when stating an experimental result R:

$$\mathrm{R} = x \pm \Delta x \qquad (1)$$

where $x$ is the expected value (single or average) of a measurement result of any physical measurable dimensional quantity/entity, and $\Delta x$ (>0) is the measurement error (accuracy).

It is not within the scope of this paper to discuss the theory of measurement error. At this point I would like to mention that errors can be improved (reduced) by increasing the number of measurements, or, by improving the measurement methods. However, even if measurement errors (accuracy) of measurable quantities do improve in the course of time, they have never become exactly zero, because of finiteness of resolution of measurement devices and because of finiteness of possible number of measurements.

It would be safe to assume that errors will never become exactly zero (same as saying that "c"— the speed of light in vacuum—is never going to change). This is basically because non-zero accuracy is a consequence of the finiteness of human capabilities, and this is a rather permanent feature of human nature.

The summary of the discussion above is that accuracy (error) of any physical measurable dimensional quantity is always finite (non-zero):

$$\Delta x \geq \varepsilon > 0 \qquad (2)$$

$\varepsilon$ is the lower limit value of error for the physical entity called x. In this section (section 1), $\varepsilon$ is positive.

Usually there is a finite ratio between a measurement result and its error $\Delta x$. It follows that measurement results of any physical measurable dimensional quantity cannot be exactly zero, i.e. they are finite. Even if $x = 0$ (which might happen), $|R| = \Delta x \geq \varepsilon > 0$. Therefore the result R can never be $R = 0 \pm 0$. Therefore:

$$|R| \geq \varepsilon > 0 \qquad (3)$$

According to the definition of physics, as stated above, we conclude that values of exactly zero should not be allowed as acceptable solutions of mathematical expressions describing legitimate physical laws, because exactly zero experimental results ($0 \pm 0$) do not occur. A corollary would be that physical laws should include $\varepsilon$ in their formulation.

Next, we discuss very shortly the infinity problem. For reasons similar to those stated above, measurement results are always finite, this time in the sense of being non-infinite:

$$|R| \leq E < \infty \qquad (4)$$

$E$ is the upper limit value for the physical entity called $x$. To summarize our discussion till now, measurement results are always finite, namely non-zero and non-infinite:

$$0 < \varepsilon \leq |R| \leq E < \infty \qquad (5)$$

Physical laws are formulated by means of mathematical expressions. As a consequence of the previous discussion, I propose that a new basic principle (postulate) of physics should be formulated, as follows:

"A legitimate law of physics is one whose mathematical expression does not allow exactly zero or infinite values as possible solutions for the measurable dimensional physical quantities present in the expression."

I call this the "Principle of Finiteness". By logical terminology, this is established as a necessary condition for the "legitimacy" of physical laws.

In the following section, we explore some implications of the finiteness principle when applied to several well-known laws of physics. Deliberately, I do this in a mostly deductive way, independently of any other physical theory or principle, besides, perhaps, the "simplicity" principle, as we shall see later.

## 2. Some Consequences of the Finiteness Principle

### 2.1- General Consequences

There are some immediate general consequences that can be derived from the finiteness principle:

a) Every measurable dimensional physical entity (quantity) has a minimum finite (non-zero) value ( $\varepsilon$ ) and a maximum finite (non-infinite) value ( $E$ ).

The minimum finite (non-zero)—or—the maximum finite (non-infinite) values of any measurable dimensional physical quantity are by definition impenetrable (irreducible) limits, namely no lower, or respectively no higher, values, are allowed.

b) Mathematical expressions of physical laws, which do not comply with the finiteness principle, should be changed accordingly in the simplest possible way (the "simplicity" principle), desirably without violating existing well-established principles.

### 2.2-Consequences of the Finiteness Principle in Special Relativity

I follow Einstein's derivation of the Lorentz Transformation [11]. We use two coordinate systems, $k$ and $k'$, whose $x$-axes permanently coincide. We consider only events on the $x$-axis. An event is represented in system $k$ by distance $x$ (from $k$-origin) and time $t$, and in system $k'$ by distance $x'$ (from $k'$-origin) and time $t'$.

The origin of $k'$ is moving relative to $k$ with velocity $v$, along the $x$-axis. We assume that at $t = t' = 0$, the origins of $k$ and $k'$ coincide, and a light-signal is transmitted from the origin along the positive $x$-axis.

We follow faithfully Einstein's derivation, with the difference that I include in my derivation a minimum finite length error $\varepsilon$ in his initial equations, as required by the Finiteness Principle ( $\varepsilon$ is the "Gedanken" minimum length error).

In this section (2.2), we assume that $\varepsilon$ is a real invariant number (same in $k$ and $k'$).

Einstein's original equations for the light signal are:

$$\left.\begin{aligned} x - ct &= 0 \\ x' - ct' &= 0 \end{aligned}\right\} \qquad \text{(for the positive } x\text{-axis)}$$

My corrected equations become:

$$\left.\begin{aligned} x - ct &= \varepsilon \\ x' - ct' &= \varepsilon \end{aligned}\right\} \qquad (6)$$

Similar relations are obtained for the negative $x$-axis.

After some calculations we get the corrected Lorentz Transformations:

$$\begin{Bmatrix} x' = \gamma\, x - v\gamma\, t + \varepsilon(1-\gamma) \\ t' = -\dfrac{v\gamma}{c^2}x + \gamma\, t + \dfrac{\varepsilon v\gamma}{c^2} \end{Bmatrix} \tag{7}$$

$$\gamma = \frac{1}{\sqrt{1 - v^2/c^2}}\,. \tag{8}$$

It is important to notice that the length measurement error $\varepsilon$ is an integral part of the Lorentz Transformations.

The corrected relativistic length contraction factor $\Gamma_{CL}$ becomes:

$$\Gamma_{CL} \equiv \frac{L_C}{L_0} = \frac{1}{\gamma}\left[1 - \frac{\varepsilon(1-\gamma)}{L_0}\right], \tag{9}$$

where $\frac{1}{\gamma}$ is the usual contraction factor ($\varepsilon = 0$), $L_0$ is length measured at rest, and $L_C$ is contracted length (should be $\geq \varepsilon$ according to the Finiteness Principle, where $\varepsilon$ is the minimum length).

Two remarks are in order:

a) The maximal possible contraction of <u>any</u> distance $L_0$ (from (9)) is ($\varepsilon / L_0$), as it should be, as required by the finiteness principle.

b) If $L_0 = \varepsilon$, $\Gamma_{CL} = 1$ (as it should be).

Next, the corrected relativistic time-dilation factor is:

$$\Gamma_{CT} \equiv \frac{\tau_D}{\tau_0} = \gamma\left[1 - \frac{v\varepsilon}{\tau_0 c^2}\right] \tag{10}$$

$\tau_0 =$ time duration measured at rest

$\tau_D =$ dilated time duration

$\gamma =$ usual time dilation ($\varepsilon = 0$)

Two remarks are in order:

a) one can test (10) by trying to detect changes in lifetimes of ultra-high velocity ($v \sim c$) and/or very short-lived subatomic particles. In this case we may use:

$$\frac{\tau_D}{\tau_0} \cong \gamma\left[1 - \frac{\varepsilon}{\tau_0 c}\right] \tag{10'}$$

to detect departure from $\gamma$ which is larger than $\gamma$'s measurement error.

b) We notice that $\Gamma_{CT}$ goes to infinity when $v \to c$. In order to satisfy the requirement of the finiteness principle, we should replace $\gamma$ by a corrected function $\gamma_C$, that stays finite for $v \to c$. It is easy to show that a simple function satisfying this requirement is:

$$\gamma_C \equiv \gamma \left[ \frac{1 - e^{-\frac{\alpha}{\gamma}}}{1 - e^{-\alpha}} \right] \qquad (11)$$

where $\alpha$ is a large positive dimensionless number, to be estimated later in this paper for sub-atomic particles. In this case $\gamma_C$ is practically identical to $\gamma$, unless $v$ is extremely close to $c$ (ultra-high velocities). In this case equation (10) becomes:

$$\frac{\tau_D}{\tau_0} = \gamma \left[ \frac{1 - e^{-\frac{\alpha}{\gamma}}}{1 - e^{-\alpha}} \right] \left[ 1 - \frac{v\varepsilon}{\tau_0 c^2} \right] \qquad (12)$$

When $\gamma >> \alpha$, we obtain (ultra-high velocities regime):

$$\frac{\tau_D{}^{\max}}{\tau_0} = \left[ \frac{\alpha}{1 - e^{-\alpha}} \right] \left[ 1 - \frac{\varepsilon}{\tau_0 c} \right] \qquad (13)$$

Since $\alpha$ is very large:

$$\frac{\tau_D{}^{\max}}{\tau_0} \approx \alpha \left[ 1 - \frac{\varepsilon}{\tau_0 c} \right] \qquad (14)$$

$\tau_D{}^{\max}$ = maximum allowable time dilation (!)

Assuming $\tau_0 >> \frac{\varepsilon}{c}$, we get:

$$\tau_D{}^{\max} \approx \alpha \tau_0 \qquad (14')$$

This and eq. (14) may be tested under appropriate conditions.

We now use (11) to correct Einstein's mass/energy equation such as to obey the Finiteness Principle.

Einstein's equation:

$$m = m_0 \gamma \qquad (15)$$

$m_0$ = rest mass

The corrected equation becomes:

$$m = m_0 \gamma \left[ \frac{1 - e^{-\frac{\alpha}{\gamma}}}{1 - e^{-\alpha}} \right] \qquad (16)$$

Then, for $\gamma >> \alpha$, we obtain:

$$m_{\max} \approx m_0 \alpha \qquad (17)$$

$m_{\max}$ = maximum attainable relativistic mass (energy) (!)

In Section 2.3 we estimate $\alpha$ for sub-atomic particles.

As we shall see, $m_{\max}$ is a fixed value for all sub-atomic particles, irrespective of $m_0$. In this case, the maximum possible relativistic mass/energy for all sub-atomic particles is a constant:

$$E_{\max} = m_{\max} c^2 \qquad (17')$$

We conclude that length contraction and time dilation depend on $\varepsilon$, $\gamma$, and also on rest values of length and time interval being measured. Finally it is easy to show that the relativistic velocity addition law and $E = mc^2$ do not change, because the new translational terms in (7) are time-independent.

### 2.3-Consequences in Quantum Mechanics

I choose to apply the Finiteness Principle to the uncertainty principle, because of its central significance in Quantum Mechanics.:

$$\Delta x \cdot \Delta p \sim h \qquad (18)$$

$$\therefore \; \Delta x \sim \frac{h}{\Delta p} \qquad (19)$$

$h$ is Planck's constant

In this discussion, $\Delta x$ and $\Delta p$ are assumed positive.

According to the Finiteness Principle, we have to correct the above equation such that $\Delta x$ cannot become less than $\varepsilon$, the minimum length error (same $\varepsilon$ as in Section 2.2) , for any finite value of $\Delta p$.

Note: in this section, (2.3), $\varepsilon > 0$.

Equation (19) represents $\Delta x$ as a decreasing function of $\Delta p$. According to the Finiteness Principle, $\Delta x$ should become a corrected function of $\Delta p$, which has a minimum finite value $\varepsilon > 0$, for a finite value of $\Delta p$.

The simplest (increasing) function to ensure this and to be added to (19) is: $a\Delta p$, where $a$ is a positive number, to be estimated in the following.

Therefore the necessary corrected equation is:

$$\Delta x \sim \frac{h}{\Delta p} + a\Delta p \qquad (20)$$

To find $a$, we set the derivative of (20) to zero, which gives the minimum:

$$-\frac{h}{(\Delta p)^2{}_{\min}} + a = 0 \qquad (21)$$

So:

$$(\Delta p)_{\min} = \sqrt{\frac{h}{a}} \qquad (22)$$

$(\Delta p)_{\min}$ is the $\Delta p$ value for which $\Delta x$ is a minimum ($\Delta x = \varepsilon > 0$).

We insert (22) in (20), and remember that: $(\Delta x)_{\min} = \varepsilon > 0$. We get:

$$\varepsilon = \frac{h}{\sqrt{\frac{h}{a}}} + a\sqrt{\frac{h}{a}} \qquad (23)$$

Then we get:

$$a = \frac{\varepsilon^2}{4h} \qquad (24)$$

Therefore the simplest <u>corrected</u> form of the uncertainty principle that <u>obeys</u> the Finiteness Principle is:

$$\Delta x \sim \frac{h}{\Delta p} + \frac{\varepsilon^2}{4h}\Delta p \qquad (25)$$

We summarize the minimum coordinates of (25):

$$\left.\begin{aligned}(\Delta x)_{\min} &= \varepsilon \\ (\Delta p)_{\min} &= \frac{2h}{\varepsilon}\end{aligned}\right\} \qquad (26)$$

Now we estimate $\alpha$ (see (11)) for particles obeying the corrected uncertainty principle. We assume that $(\Delta p)_{\min}$ is also the maximum value of $\Delta p$ if we want equation (25) to be a single-valued monotonous function.

We obtain:

$$\frac{2h}{\varepsilon} = \max(\gamma_C\, m_0 v) \qquad (27)$$

(since $\gamma_C$ replaces $\gamma$ for $v \sim c$, see equation (11)); therefore:

$$\max(\gamma_C\, m_0 v) = m_0 \max(\gamma_C\, v) = m_0 \max(\gamma_C\, c) = m_0 c \max(\gamma_C) = m_0 c\alpha$$

Therefore:

$$\frac{2h}{\varepsilon} = \alpha\, m_0 c$$

$$\therefore \qquad \alpha = \left(\frac{2h}{\varepsilon\, c}\right)\frac{1}{m_0} \tag{28}$$

Then equation (17) becomes:

$$m_{\max} = m_0 \alpha = \frac{2h}{\varepsilon c} \tag{29}$$

Equation (17') becomes:

$$E_{\max} = \alpha\, m_0 c^2 = \frac{2h}{\varepsilon\, c} c^2 = \frac{2hc}{\varepsilon} \tag{30}$$

Therefore we conclude that all sub-atomic particles have the same maximum possible relativistic mass/energy, irrespective of their rest mass/energy. They obtain this value at $v \approx c$ (ultra-high velocities).

## 3. Discussion

Before continuing, we remark parenthetically that contradictions may arise between the consequences of the finiteness principle and the consequences of other well-established principles. In this case, the discrepancies might occur because different principles may deal with different parameters regime (e.g., different velocity ranges).

Returning to the discussion on the results obtained in this paper, we first note that equation (25), with $\varepsilon = l_p$ (Planck's length) is required in quantum gravity as a replacement for the usual uncertainty principle, and it is referred to as the "generalized uncertainty principle" [12]. In this context a generalized quantum-mechanical Hamiltonian is constructed, on the basis of the generalized uncertainty principle, which becomes important at Planck scale. In Ref. [13] a generalized uncertainty principle is derived from micro-black hole Gedanken experiment, using only the uncertainty principle and $R_s$ (Schwarzschild radius).

Historically, a generalized uncertainty principle was first derived in [14] and then in [15] with (our) $\varepsilon = l_p$, and $l_p^2 \approx \alpha'$ (string tension).

From the results obtained in our paper, it is clear that the Finiteness Principle explains in a fundamental way the additional term in the generalized uncertainty principle ($\varepsilon = l_p$ is a special case, representing the inevitable gravitational interference uncertainty). It is indeed very encouraging to find that quantum gravity complies with the Finiteness Principle.

It is also important to mention that the T-Duality in string theory requires a minimum observable length $R_{\min} \sim \alpha'^2$ and brings about a generalized uncertainty principle [16, 17]. This means that T-Duality (and maybe other aspects of string theory) is intimately connected with the Finiteness Principle.

Next, we note that neutrino measurements data are very important in testing the consequences of the Finiteness Principle, because of their $m_0 \neq 0$, and $v \sim c$ [see eqs. (16), (17), (17'), (29), (30), with $\varepsilon = l_p$ (Planck length) and $m_{\max} = m_p$ (Planck mass)]. It is especially significant to have experiments giving each at least two of the three following parameters: energy, rest-mass, speed.

However, the usefulness of the existing neutrino data [18, 19, 20, 21, 22, 23] is limited, because of lack of ultra-high energy measurements, lack of absolute mass measurements, and finally, lack of sufficient accuracy of speed measurement at $v \sim c$.

Finally, because of the importance of the fundamental constants to the Planck scale limits, astronomical measurements are being made and evaluated extensively, in order to determine the accuracy and constancy of various parameters [24].

## 4. Conclusions

I propose a new postulate in physics, the Principle of Finiteness. I believe it is a very basic postulate, to which any "legitimate" physical law should obey (comply). I explore the implications of the finiteness principle to some simple relations of special relativity and quantum mechanics.

The corrected relativistic length contraction and time dilation factors depend on: $\varepsilon$, $\gamma$, $\tau_0$, $L_0$, and $m_0$ (through $\alpha$).

In general, when corrected to comply with the Finiteness Principle, physical laws should include the minimum measurement error $\varepsilon$ in their mathematical formulation.

An important prediction is that the maximum relativistic mass/energy attainable is the same for all sub-atomic particles (independent of $m_0$).

A possible way to interpret this result would be in terms of Planck's units, as follows [25, 26, 27, 28]: at $v \cong c$, all subatomic particles shrink to $\varepsilon = l_p$ (Planck's length), achieve a relativistic maximum mass of $m_{\max} \equiv m_p$ (Planck's mass), become micro-black-holes, and disappear in a flash of Hawking's radiation consisting of gamma-rays having a cut-off energy of $m_p / 2$ at the appropriate Compton minimum wavelength. This would mean that at $v \cong c$, mass undergoes a "phase transition" into pure radiant energy, satisfying the demand of the Standard Model that mass is ≈ zero at $v \cong c$ (maybe leaving minimal remnants).

It also seems that the generalized uncertainty principle required by quantum gravity (and T-duality) as a replacement for the usual Heisenberg uncertainty principle, is actually a necessary consequence of the Finiteness Principle. It follows that quantum gravity complies with the Finiteness Principle. The Finiteness Principle apparently explains in a fundamental way the need of quantum gravity to replace the usual uncertainty principle with a corrected ("generalized") version, at the minimum length scale.

Therefore, the Finiteness Principle may contribute to the axiomatic foundation of quantum gravity.

## Acknowledgements

It is my pleasure to acknowledge useful remarks by: Prof. J. Bekenstein, Prof. L. Dorman, Prof. N. Fisch, Prof. S. Cohen, Prof. A. Yahalom, Prof. L. Horwitz, Dr. A. Yarom, and Prof. A. Polyakov.

I would like to thank J. A. Baumgaertel and L. Peterson for carefully preparing the manuscript for publication.